\documentclass[12pt]{iopart}
\usepackage{graphicx}
\usepackage{dcolumn}
\usepackage{bm}
\usepackage{epsf}

\begin{document}

\title{Excitation transfer in two two-level systems coupled to an oscillator}
\author{P L Hagelstein$^1$, I U Chaudhary$^2$}

\address{$^1$ Research Laboratory of Electronics, 
Massachusetts Institute of Technology, 
Cambridge, MA 02139,USA}
\ead{plh@mit.edu}

\address{$^2$ Research Laboratory of Electronics, 
Massachusetts Institute of Technology, 
Cambridge, MA 02139,USA}
\ead{irfanc@mit.edu}

\begin{abstract}
We consider a generalization of the spin-boson model in which two different two-level
systems are coupled to an oscillator, under conditions where the oscillator energy
is much less than the two-level system energies, and where the oscillator is highly
excited.
We find that the two-level system transition energy is shifted, producing a Bloch-Siegert
shift in each two-level system similar to what would be obtained if the other were absent.
At resonances associated with energy exchange between a two-level system and the oscillator,
the level splitting is about the same as would be obtained in the spin-boson model at
a Bloch-Siegert resonance.
However, there occur resonances associated with the transfer of excitation between
one two-level system and the other, an effect not present in the spin-boson model.
We use a unitary transformation leading to a rotated system in which terms responsible
for the shift and splittings can be identified.
The level splittings at the anticrossings associated with both energy exchange and 
excitation transfer resonances are accounted for with simple two-state models
and degenerate perturbation theory using operators that appear in the rotated
Hamiltonian.

\end{abstract}


\maketitle

\section{Introduction}
\label{sec:intro}

The coupled quantum system consisting of a 
two-level system interacting with a harmonic oscillator
provides a model that has been focus of a large number
of studies over recent decades.
A reduced (Rabi Hamiltonian) version of the problem was published by Bloch and Siegert 
in studies of the interaction of a dynamical magnetic field perpendicular to a static
magnetic field with a spin system \cite{BlochSiegert},
and was later applied to the problem of atoms interacting with 
an electromagnetic mode \cite{Shirley}.
The more complete model (spin-boson Hamiltonian) was introduced much later
by Cohen-Tannoudji and collaborators \cite{Cohen}.

At the most basic level, the interaction of the two-level system with the oscillator
produces a shift in the two-level energy (the Bloch-Siegert shift).
Energy exchange between the two-level and oscillator is allowed when the shifted
two-level system energy becomes resonant with an odd number of oscillator quanta,
which results in level anticrossings at these resonances (Bloch-Siegert resonances).
If the characteristic energy of the oscillator $\hbar \omega_0$ is much less than
the transition energy $\Delta E$, then many oscillator quanta are required to
match the shifted (dressed) transition energy.
In such a limit, the model is considered to be in the multiphoton regime;
which is a topic of current interest \cite{Fregenal,Forre,OstrovskyHorsdal}.
The models under consideration in this work are studied in the multiphoton regime.

If the oscillator is highly excited, then the problem simplifies.
The two-level system interacts with the oscillator to produce a shift as before; however, 
the oscillator is only weakly impacted by the two-level system since the two-level
energy in this case amounts to a small fraction of the total oscillator energy.
In this limit the Bloch-Siegert shift is accurately approximated using an adiabatic model \cite{OstrovskyHorsdal}, 
and we have obtained new estimates for the level splitting at the anticrossings \cite{HagelsteinChau2} using degenerate perturbation
theory on a rotated version of the model.
In the rotated problem, multiphoton transitions are mediated through a complicated
``perturbation'' operator, but in the end the level-splitting is obtained approximately
from a simple two-state model with only first-order coupling.
The complicated interactions of the original spin-boson problem, which are responsible for
energy exchange of a large number of quanta, are reasonably well accounted for through
the lowest-order coupling in the rotated version of the problem.

This approach, and the resulting conceptual simplification of the problem, is reasonably general
and very powerful.
We extended the analysis to the case of a spin-one system coupled to an oscillator \cite{HagelsteinChau3}, 
and obtained results for the Bloch-Siegert shift in agreement with previous work, as well as
new results for the level splitting at the Bloch-Siegert resonances.
The accuracy was comparable to that obtained for the spin-boson model, and it seems clear that the
approach can be applied systematically to higher-spin generalizations of the spin-boson model
as well.
We also studied a different generalization of the spin-boson model in which a three-level system
is coupled to an oscillator \cite{HagelsteinChau4}.
Technical issues associated with the three-level system made the implementation of the rotation
much more challenging; however, in the end we obtained good results for the level shifts, and
for the level splitting at the anticrossings, as long as the anticrossing occured away from
other strong resonances which interfere.

In this work we turn our attention to a different generalization of the spin-boson model in which
two different two-level systems are coupled to an oscillator.
This and related models have been studied in connection with studies of the interaction between
two atoms and a cavity \cite{Mahmood1987,Iqbal1988,Jex1990}, and quantum entanglement \cite{Mancini2001,Brennan2000,Jessen2001}.
Our approach is most useful in the multiphoton regime with a highly excited oscillator.
In this case, both two-level systems experience Bloch-Siegert shifts due to their interaction
with the oscillator, and anticrossings occur associated with energy exchange between each
two-level system and the oscillator.
The Bloch-Siegert shift for each two-level system in this case is very close to what would be
expected if the other two-level system were absent, and the level splittings at the energy
exchange anticrossings are not very different from the single two-level version of the problem.
What is new in this problem are anticrossings associated with excitation transfer, in which
the excitation from one two-level system is transferred to the other.
In the rotated version of this model, we find a spin-spin interaction term which mediates
these new excitation transfer transitions.
Our attention in this work is then focused on excitation transfer, and we find that a simple
two-state model and degenerate perturbation theory leads to reasonably accurate estimates
of the level splittings away from other resonances.

The excitation transfer effect in this model is very weak.  
To study it in a regime in which energy exchange resonances do not interfere, we need to work
with low oscillator energy (which maximizes the number of oscillator quanta needed for resonance)
and modest oscillator excitation (since the level splitting is inversely proportional
to the number of oscillator quanta).
The reason why excitation transfer is so weak in this model is studied using a finite-basis approximation;
in which we see that destructive interference occurs between the contributions from pathways involving
intermediate states.

We briefly examine the possibility of reducing or eliminating the destructive interference,
in order to make the excitation transfer effect stronger.
If the model is modified so that the coupling between the different two-level systems
is through conjugate oscillator operators, then some of the destructive interference is removed.
If the model is augmented with loss terms that remove the contribution of the lower-energy intermediate
states, then the destructive interference is completely eliminated; the excitation transfer
effect then becomes much stronger.

\newpage
\section{Model}

We consider the model described by the Hamiltonian

{\small

\begin{equation}
\label{eq:TwoTwoLevelHamiltonian}
\hat{H} 
~=~
\Delta E_1 {\hat{s}_z^{(1)} \over \hbar}
+
\Delta E_2 {\hat{s}_z^{(2)} \over \hbar}
+
\hbar \omega_0  \hat{a}^\dagger \hat{a} 
+
U_1 (\hat{a}^\dagger+\hat{a}) {2 \hat{s}_x^{(1)} \over \hbar}
+
U_2 (\hat{a}^\dagger+\hat{a}) {2 \hat{s}_x^{(2)} \over \hbar}
\end{equation}

}

\noindent
There are a pair of two-level systems, with unperturbed transition energies $\Delta E_1$ and $\Delta E_2$;
and an oscillator, with characteristic energy $\hbar \omega_0$.
The first two-level system interacts linearly with the oscillator, with a coupling strength of $U_1$; 
similarly, the second two-level system also interacts linearly with the oscillator, with a coupling strength of $U_2$.
The spin operators are defined in terms of the Pauli matrices according to

\begin{equation}
\hat{\bf s}^{(j)} 
~=~  
{\hbar  \over 2} \hat{\boldsymbol{\sigma}}^{(j)}
\end{equation}

\noindent
where the superscript denotes which two-level system is referenced.
In the multiphoton regime, the characteristic energy of the oscillator is much less than the
unperturbed transition energies

\begin{equation}
\hbar \omega_0 ~\ll~ \Delta E_1
\ \ \ \ \ \ \ \ \
\hbar \omega_0 ~\ll~ \Delta E_2
\end{equation}

\newpage
\section{Unitary transformation}

As in our previous work on the spin-boson problem \cite{HagelsteinChau2,HagelsteinChau1}, 
we find it useful to consider a unitary equivalent Hamiltonian

\begin{equation}
\hat{H}^\prime ~=~ \hat{\mathcal{U}} \hat{H} \hat{\mathcal{U}}^\dagger 
\end{equation}
\noindent where

\begin{equation}
\hat{\mathcal{U}}
~=~
\hat{\mathcal{U}}_1 \hat{\mathcal{U}}_2
~=~
e^{-i \lambda_1 \hat{\sigma}_y^{(1)} }
e^{-i \lambda_2 \hat{\sigma}_y^{(2)} }
\end{equation}

\noindent 
Since $\boldsymbol{\sigma}^{(1)}$ commutes with $\boldsymbol{\sigma}^{(2)}$, 
the computation is very similar to the single two-level problem. 
This unitary transform is a straightforward generalization of one used previously
for the spin-boson problem \cite{Wagner,LarsonStenholm}

\subsection{Dressed Hamiltonian and ``unperturbed'' part}

We can write the rotated Hamiltonian as

\begin{equation}
\hat{H}^\prime
~=~
\hat{H}_0
+
\hat{V}_1
+
\hat{V}_2
+
\hat{W}_1
+
\hat{W}_2
+
\hat{V}_{12}
\end{equation}

\noindent 
where the ``unperturbed'' part of the rotated Hamiltonian $\hat{H}'$ is given by 
 
\begin{equation}
\hat{H}_0
~=~
\sqrt{ \Delta E_1^2 + 8U_1^2  y^2 }
{\hat{s}^{(1)}_z \over \hbar}
+
\sqrt{ \Delta E_2^2 + 8U_2 y^2 }
{\hat{s}^{(2)}_z \over \hbar}
+
\hbar \omega_0 \hat{a}^\dagger \hat{a} 
\end{equation}

\noindent
where

\begin{equation}
y ~=~ {\hat{a} + \hat{a}^\dagger \over \sqrt{2}}
\end{equation}

\noindent
This is similar to the unperturbed part that we obtained previously in the case of the spin-boson
model, where now two dressed two-level terms appear instead of one.

In the spin-boson model, and also in other problems that we have studied, the ``unperturbed'' part $\hat{H}_0$
of the rotated Hamiltonian $\hat{H}'$ results in a good approximation for the oscillator and
dressed transition energy of the two-level systems.
It can be used to develop approximations for resonance conditions as we discuss in the next section.
Because of this, we have come to think of $\hat{H}_0$ as describing an unperturbed version of
the dressed problem in which no interactions occur at level crossings.
Viewed in this way, the other terms in the rotated Hamiltonian can be thought of as perturbations.

\subsection{Perturbations involved in energy exchange resonances}

There are now two primary perturbations $\hat{V}_1$ and $\hat{V}_2$ (only a single one
appears in the spin-boson problem since there is only one two-level system in that model), 
each of which can be described by the general formula

{\small

\begin{equation}
\hat{V}_j
~=~
i {\hbar \omega_0 \over 2}
\left \lbrace
{
\left ( 
\displaystyle {\sqrt{2} U_j  \over \Delta E_j }
\right )
\over 
\left ( 1 + \displaystyle{8U_j^2 y^2 \over \Delta E_j^2} \right )
}
{d \over dy} 
+
{d \over dy} 
{
\left (
\displaystyle {\sqrt{2} U_j  \over \Delta E_j }
\right )
\over 
\left ( 1 + \displaystyle{8U_j^2 y^2 \over \Delta E_j^2} \right )
}
\right \rbrace
{2\hat{s}^{(j)}_y \over \hbar}
\end{equation}

}

\noindent
where

\begin{equation}
{d \over dy} ~=~ {\hat{a} - \hat{a}^\dagger \over \sqrt{2}}
\end{equation}

\noindent
In the multiphoton regime, these terms give rise to level splittings at
resonances in which one unit of excitation of a two-level system is
exchanged for an odd number of oscillator quanta, as discussed in
\cite{HagelsteinChau2}.

\subsection{Potential operators}

There are two small terms $\hat{W}_1$ and $\hat{W}_2$, both described through
the general formula

{\small

\begin{equation}
\hat{W}_j
~=~
\hbar \omega_0 
{
\left (
\displaystyle {U_j  \over \Delta E_j }
\right )^2
\over 
\left ( 1 + \displaystyle{8U_j^2 y^2 \over \Delta E_j^2} \right )^2
}
\left ( {2 \hat{s}^{(j)}_y \over \hbar} \right )^2
\end{equation}

}

\noindent
Since the square of the spin operator $\hat{s}_y$ is proportional
to the identity matrix

\begin{equation}
\left ( {2 \hat{s}_y \over \hbar} \right )^2
~=~
\hat{\sigma}_y^2
~=~
\left (
\begin{array} {cc}
1 & 0 \cr
0 & 1 \cr
\end{array}
\right )
\end{equation}

\noindent
these terms become simple potentials

{\small

\begin{equation}
\hat{W}_j
~=~
\hbar \omega_0 
{
\left ( 
\displaystyle {U_j  \over \Delta E_j }
\right )^2
\over 
\left ( 1 + \displaystyle{8U_j^2 y^2 \over \Delta E_j^2} \right )^2
}
\end{equation}

}

\noindent
In the large $n$ limit which is of interest in this paper, these potentials
are very small, and do not contribute in a significant way to either
the occurrence of resonances or the level splitting at resonance.  
We neglect them in what follows.

\subsection{Spin-spin interaction}

Finally, we find a spin-spin operator $\hat{V}_{12}$ given by

\vskip 0.1in

{\small

\begin{equation}
\hat{V}_{12}
~=~
2\hbar \omega_0 
{
\left ( 
\displaystyle {U_1  \over \Delta E_1 }
\right )
\over 
\left ( 1 + \displaystyle{8U_1^2 y^2 \over \Delta E_1^2} \right )
}
{
\left ( 
\displaystyle {U_2 \over \Delta E_2 }
\right )
\over 
\left ( 1 + \displaystyle{8U_2^2y^2 \over \Delta E_2^2} \right )
}
\left ( {2 \hat{s}^{(1)}_y \over \hbar} \right )
\left ( {2 \hat{s}^{(2)}_y \over \hbar} \right )
\end{equation}

}

\vskip 0.1in

\noindent
This operator has no analog in the spin-boson model.
In the multiphoton regime this term contributes to the level
splitting associated with resonances involving excitation transfer
(where one unit of excitation in one two-level system is exchanged
for one unit of excitation in the other two-level system, accompanied
by the exchange of an even number of oscillator quanta).
As this effect is new in this model, it will be the focus of our
attention in this work.

\newpage

\section{Energy levels and resonance conditions}

The time-independent Schr\"odinger equation for $\hat{H}_0$ is

{\small

$$
\left ( E + {\hbar \omega_0 \over 2} \right ) \Psi
~=~
{\hbar \omega_0  \over 2} \left ( - {d^2 \over dy^2} + y^2 \right ) \Psi 
+
\sqrt{ \Delta E_1^2 + 8U_1^2  y^2 }
{\hat{s}^{(1)}_z \over \hbar}\Psi
\ \ \ \ \ \ \ \ \ \ \ \ \ \ \ \ \ \ \ \ \
\ \ \ \ \ \ \ \ \ \ \ \ \ \ \ 
$$
\begin{equation}
\ \ \ \ \ \ \ \ \ \ \ \ \ \ \ \ \ \ \ \ \
\ \ \ \ \ \ \ \ \ \ \ \ \ 
+
\sqrt{ \Delta E_2^2 + 8U_2 y^2 }
{\hat{s}^{(2)}_z \over \hbar}\Psi
\end{equation}
}

\subsection{Product solutions}

We can reduce this Schr\"odinger equation to a purely spatial problem in $y$ 
by assuming a product wavefunction for $\Psi$ of the form

\begin{equation}
\Psi 
~=~
u_{m_1,m_2}(y) |s,m_1 \rangle |s,m_2 \rangle
\end{equation}

\noindent 
where the $|s,m_j\rangle$ are spin 1/2 eigenkets (with $s = 1/2$ and $m_j = \pm 1/2$).  
This leads to a one-dimensional Schr\"odinger equation 

\begin{equation}
\epsilon u(y)
~=~
\left [
- {d^2 \over dy^2} 
+v(y)
\right ] u(y)
\label{Schrod_y}
\end{equation}

\noindent
where the normalized energy eigenvalue $\epsilon$ is

\begin{equation}
\epsilon 
~=~ 
{2 E \over \hbar \omega_0}
+
1
\end{equation}

\noindent
and where the normalized nonlinear potential $v(y)$ is

\begin{equation}
v(y)
~=~
y^2
+
{2 m_1 \over \hbar \omega_0}
\sqrt{ \Delta E_1^2 + 8U_1^2  y^2 }
+
{2 m_2 \over \hbar \omega_0}
\sqrt{ \Delta E_2^2 + 8U_2^2  y^2 }
\end{equation}

\noindent 
We have suppressed the subscripts $m_i$ on $u(y)$.

\subsection{Approximate solutions}

In the large $n$ limit, the oscillator is highly excited and we would not expect
the two-level systems to have much impact on the oscillator.  
In this case, a reasonable approximation for the oscillator is to use eigenfunctions
of the simple harmonic oscillator

\begin{equation}
u(y) ~=~ \langle y|n \rangle
\end{equation}

\noindent 
where $| n \rangle$ are the harmonic oscillator eigenfunctions.
Within this approximation, we may write

\begin{equation}
E_{n,m_1,m_2}
~=~
\Delta E_1(g_1) m_1 
+
\Delta E_2(g_2) m_1 
+
n \hbar \omega_0 
\end{equation}

\noindent
where the dressed transition energies $\Delta E_1(g_1)$ and $\Delta E_2(g_2)$ are
given by

\begin{equation}
\Delta E_j(g_j) 
~=~ 
\Delta E_j 
\left \langle n \left |
\sqrt{ 1 + {8U_j^2  y^2 \over \Delta E_j^2 }}
\right | n  \right \rangle
\end{equation}

\noindent
The dimensionless coupling constants $g_1$ and $g_2$ are defined according to

\begin{equation}
g_j ~=~ \frac{U_j \sqrt{n}}{\Delta E_j}
\end{equation}

We interpret $\Delta E_j(g_j)$ as the ``dressed'' transition energy of
the $j$th two-level system.  
Numerical solutions for the original Hamiltonian $\hat{H}$ are well described by this expression 
away from resonances, similar to what we found in the spin-boson problem \cite{HagelsteinChau2}.
This also allows us to meaningfully label our eigenfunctions $u(y)$ with $n$ 
(again suppressing the weak $m_1$ and $m_2$ dependence)

\begin{equation}
u(y) = u_{n}(y)
\end{equation}

\subsection{Energy exchange resonances}

\begin{figure} [t]
\begin{center}
\mbox{\epsfbox{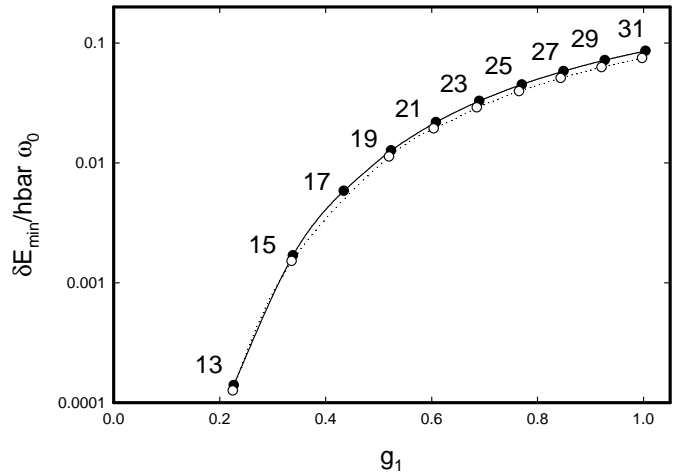}}
\caption{Level splittings associated with energy exchange anticrossings, for a model
with $\Delta E_1 = 11 ~\hbar \omega_0$ and
$\Delta E_2 = ~13 \hbar \omega_0$, $n_0 = 10^4$.
Solid circles -- results of direct numerical calculation of the original $\hat{H}$ problem; 
open circles  -- results from degenerate perturbation theory using eigenfunctions of
$\hat{H}_0$. 
The dimensionless coupling coefficient $g_2$ is fixed at 0.5 for these calculations.
The number of oscillator quanta exchanged $\Delta n$ is indicated for each anticrossing.
}
\label{dEminvsg1}
\end{center}
\end{figure}

Within this approximation scheme, we can develop resonance conditions for energy exchange
resonances.
For energy exchange between the first two-level system and the oscillator, the resonance
condition is

\begin{equation}
\Delta E_1(g_1) ~=~ \Delta n \hbar \omega_0
\end{equation}

\noindent
with $\Delta n$ odd. 
These resonances correspond to the ones we analyzed using a similar approach in the
spin-boson model \cite{HagelsteinChau2}.
The level splittings at the associated anticrossings are illustrated in Figure \ref{dEminvsg1}
under conditions where the energy exchange resonances dominate (open circles).
Also shown are approximate results using degenerate perturbation theory based on
the eigenfunctions of the rotated $\hat{H}_0$ problem.
In this calculation, one state is chosen in which the first two-level system is excited,
the second is in the ground state, and $n$ quanta are present in the oscillator;
the other state is chosen so that both two-level systems are in the ground state,
and $n+\Delta n$ quanta are in the oscillator.
The level splitting at the anticrossings using degenerate perturbation theory
is  \cite{HagelsteinChau2}

\begin{equation}
\delta E_{min} ~=~ 2|\langle \Psi_{n,1/2,-1/2}|\hat{V}_1|\Psi_{n+\Delta n, -1/2,-1/2} \rangle |
\end{equation}

\noindent
One sees that the level splittings are in reasonable agreement with this approximation.
This is similar to what we found previously in the spin-boson model.
Moreover, the level splitting found in this example matches that for the equivalent
spin-boson model (obtained by removing the second two-level system) such
that they could not be distinguished if plotted together in this figure.

Energy exchange between the second two-level system and oscillator occurs similarly when
the resonance condition

\begin{equation}
\Delta E_2(g_2) ~=~ \Delta n \hbar \omega_0
\end{equation}

\noindent
is satisfied.

\epsfxsize = 3.60in
\epsfysize = 2.60in

\subsection{Excitation transfer resonances}

This system supports another kind of resonance that is not present in the spin-boson model.
Excitation can be transferred from one two-level system to the other (along with energy
exchange with the oscillator), which has motivated us to refer to the associated
resonances as ``excitation transfer'' resonances.
Consider the resonance between one state, with energy

\begin{equation}
E_{n_0,-1/2,1/2} 
~=~ 
- {1 \over 2} \Delta E_1(g_1) 
+ {1 \over 2} \Delta E_2(g_2) 
+ \hbar \omega_0  n_0 
\end{equation}

\noindent
and another, with energy

\begin{equation}
E_{n_0+\Delta n,1/2,-1/2} 
~=~ 
  {1 \over 2} \Delta E(g_1) 
- {1 \over 2} \Delta E(g_2) 
+ \hbar \omega_0 (n_0+\Delta n) 
\end{equation}

\noindent
We obtain the resonance condition by requiring the two basis states to
have the same energy

\begin{equation}
E_{n_0,-1/2,1/2}
~=~
E_{n_0+\Delta n,1/2,-1/2}
\end{equation}

\noindent
This is consistent with the constraint

\begin{equation}
\Delta E(g_2) - \Delta E(g_1) ~=~ \Delta n \hbar \omega_0
\label{ResCond}
\end{equation}

\noindent
with $\Delta n$ even.

\epsfxsize = 3.70in
\epsfysize = 2.90in
\begin{figure} [t]
\begin{center}
\mbox{\epsfbox{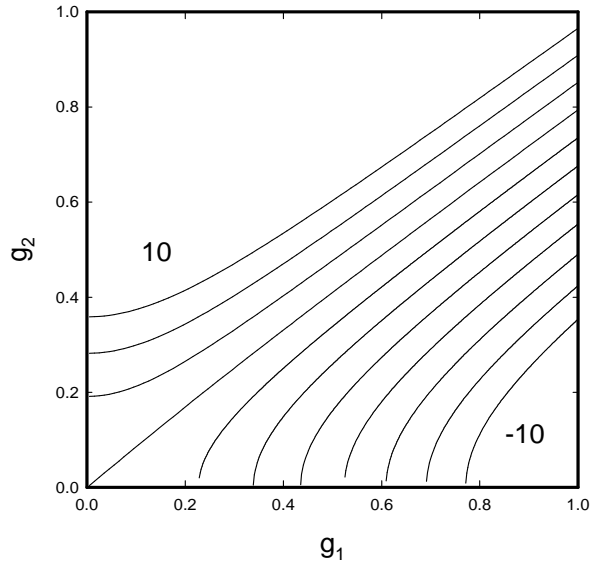}}
\caption{$\Delta E_2 = 15 \hbar \omega_0$, $n_0 = 9600$, and $\Delta n$
  ranging between 
-10 and 10. The lines associated with $\Delta n = -10$ and $\Delta n = 10$ are
denoted, with the lines in between corresponding to even $\Delta n$. } 
\label{Fig35x1}
\end{center}
\end{figure}

  Results from a computation based on the WKB approximation of resonance conditions are illustrated in Figure \ref{Fig35x1}.
The WKB approximation for this case is discussed in Appendix A.
One observes that as long as $\Delta E_2$ is greater than $\Delta E_1+\Delta n \hbar \omega_0$,
for every $g_2$ a value can be found for $g_1$ at which a resonance occurs.  
Similarly, as long as $\Delta E_1$ is greater than $\Delta E_2 - \Delta n \hbar \omega_0$, for
every $g_1$ there occurs a value of $g_2$ at which a resonance occurs.

\newpage
\section{Level splittings for excitation transfer resonances}

As we observed in the case of a single two-level system, energy levels split
at anticrossing when resonances occur.  We can compute this splitting in
at least two ways: either by direct numerical diagonalization of the full
unrotated Hamiltonian $\hat{H}$, or by using degenerate perturbation theory on
the rotated Hamiltonian.

\subsection{Degenerate perturbation theory}

In the vicinity of a level anticrossing that is free of energy exchange resonance
disruption, we can investigate the level splittings using a two-state approximation
of the form

\begin{equation}
\psi
~=~
c_0 \phi_0
~+~
c_1 \phi_1
\end{equation}

\begin{equation}
\phi_0 ~=~ u_n(y) |s,-1/2 \rangle |s,1/2 \rangle
\ \ \ \ \ \ \ \ \ \
\phi_1 ~=~ u_{n+\Delta n}(y) |s,1/2 \rangle |s,-1/2 \rangle
\end{equation}

\noindent
The two-state problem leads to the following characteristic
equation for 
the energy levels

\begin{equation}
E
\left (
\begin{array} {c}
c_0 \cr
c_1 \cr
\end{array}
\right )
~=~
\left (
\begin{array} {cc}
E_0                                            & \langle \phi_0 | \hat{V}_{12} | \phi_1 \rangle \cr
\langle \phi_1 | \hat{V}_{12} | \phi_0 \rangle & E_1                                            \cr
\end{array}
\right )
\end{equation}

\noindent 
Since we are at resonance
\[
E_0 = E_1
\]
\noindent and the level-splitting is given by
given by 
\begin{equation}
\delta E_{min} = 2 |\langle \phi_0|\hat{V}_{12}|\phi_1\rangle |
\end{equation}

\subsection{Level splitting estimates}

The matrix element that appears here can be written as

\begin{equation}
\label{eq:V12}
\langle \phi_0 | \hat{V}_{12} | \phi_1 \rangle 
~=~
2\hbar \omega_0 
{U_1  \over \Delta E_1 }
{U_2 \over \Delta E_2 }
I_{12}
~=~
 {2 \hbar \omega_0 g_1 g_2  \over n}
I_{12} 
\end{equation}

\noindent
where the integral $I_{12}$ is given by

\begin{equation}
\label{eq:I12}
I_{12}
~=~
\int_{-\infty}^\infty
u_n(y)
{1 
\over
\left [ 1 + \displaystyle{8U_1^2 y^2 \over \Delta E_1^2} \right ]
\left [ 1 + \displaystyle{8U_2^2y^2 \over \Delta E_2^2} \right ]
}
u_{n+\Delta n}(y)
dy
\end{equation}

\noindent
We can compute this integral using either numerical wavefunctions obtained by diagonalizing $\hat{H}_0$, 
or by utilizing the WKB approximation \cite{HagelsteinBook}.

The level splitting on resonance can be expressed in terms of the integral $I_{12}$ as

\begin{equation}
\label{eq:DeltaEmin}
\delta E_{min} 
~=~
{4 \hbar \omega_0 g_1 g_2 \over n} I_{12} 
\end{equation}

\begin{figure} [t]
\begin{center}
\mbox{\epsfbox{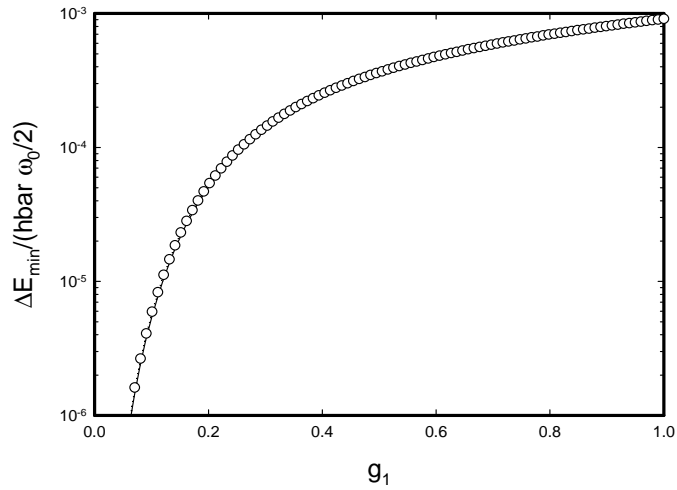}}
\caption{
Level splittings for $\Delta E_1 = 53 ~\hbar \omega_0$ and
$\Delta E_2 = 51 ~\hbar \omega_0$, $n_0 = 1000$, and $\Delta n=-2$.  
Solid circles -- results of direct numerical calculation of the original $\hat{H}$ problem; 
solid line    -- results from degenerate perturbation theory using eigenfunctions of
$\hat{H}_0$. 
The results are shown as a function of $g_1$, where $g_2$ has been optimized to minimize the level splitting. 
} 
\label{DEZvsg1}
\end{center}
\end{figure}

\noindent
The results of the level splittings from a direct numerical computation using the 
unrotated $\hat{H}$ and degenerate perturbation theory (using the WKB approximation) 
are illustrated in Figure \ref{DEZvsg1}.
We can see that the perturbation theory results are in excellent agreement with the exact result.
In this example, we have selected large two-level system transition energies relative to
the oscillator energy in order to minimize the impact of energy exchange resonances.  
In addition, we have chosen a moderate value for $n$ (instead of a larger value)
to increase the level splitting.

\epsfxsize = 3.60in
\epsfysize = 2.60in

\newpage

\section{Excitation transfer using a finite-basis expansion}

The level splitting associated with an excitation transfer resonance
is a weak effect in this model.
We can better understand the slow dynamics of the excitation transfer by 
making use of a finite basis expansion.
One can see from this kind of calculation that destructive interference
between different pathways produces a very small second-order coupling
between initial and final states.
If this destructive interference can be broken, then the
second-order coupling is greatly increased.

\subsection{Finite-basis approximation}

Indirect coupling between initial and final states associated with
an excitation transfer process can be analyzed in the weak coupling limit
through the use of a finite-basis approximation.
Consider a finite-basis approximation with six basis states 

\begin{equation}
\Psi ~=~ \sum_{j=1}^6 ~c_j~\Phi_j
\end{equation}

\noindent
where the basis states are
{\small
\begin{eqnarray}
\Phi_1 & = & |n   \rangle |s, 1/2 \rangle |s,-1/2 \rangle \nonumber \\
\Phi_2 & = & |n-1 \rangle |s,-1/2 \rangle |s,-1/2 \rangle \nonumber \\
\Phi_3 & = & |n+1 \rangle |s,-1/2 \rangle |s,-1/2 \rangle \nonumber \\
\Phi_4 & = & |n-1 \rangle |s, 1/2 \rangle |s, 1/2 \rangle \nonumber \\
\Phi_5 & = & |n+1 \rangle |s, 1/2 \rangle |s, 1/2 \rangle \nonumber \\
\Phi_6 & = & |n   \rangle |s,-1/2 \rangle |s, 1/2 \rangle 
\end{eqnarray}
}

\epsfxsize = 3.00in
\epsfysize = 2.80in
\begin{figure} [t]
\begin{center}
\mbox{\epsfbox{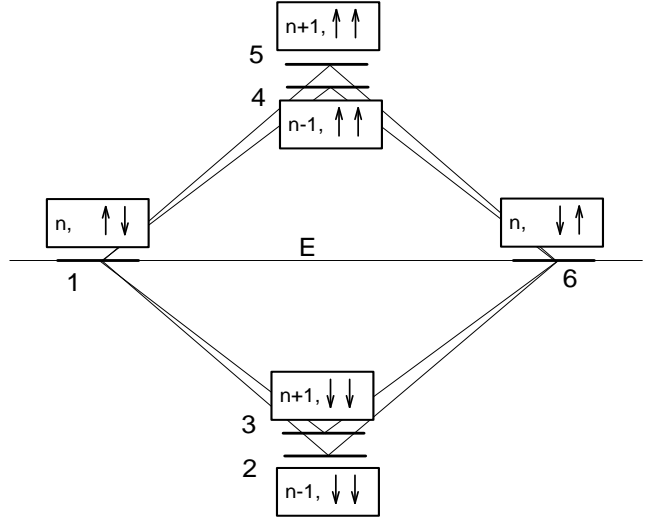}}
\caption{Schematic of levels for a pair of two-level systems with indirect coupling
between two degenerate states.
The basis state index $j$ is indicated near the energy level.
Boxes are included that give the number of oscillator quanta; arrows indicate
whether the first two-level system is excited or not (first arrow), and whether the second
two-level system is excited or not (second arrow).} 
\label{levels3}
\end{center}
\end{figure}

\noindent
The energy levels and coupling are indicated schematically in Figure \ref{levels3}.
The excitation transfer process in this case would take the system from
an initial state $\Phi_1$ (with an excited first two-level system, and a
ground state second two-level system) to a final state $\Phi_6$ (with
a ground state first two-level system, and an excited second two-level
system).
Since the Hamiltonian $\hat{H}$ does not couple the basis state
$\Phi_1$ to the basis state $\Phi_6$ directly, the coupling between
the two states is indirect.
The intermediate states $\Phi_2, \dots \Phi_5$ will provide the
dominant pathways between the initial and final states in the
case of weak coupling.

\subsection{Indirect coupling}

It is possible to obtain an indirect interaction between basis state
$\Phi_1$ and basis state $\Phi_6$ by eliminating the intermediate
states algebraically, as we illustrate in what follows.
%
%
The finite-basis equations for the expansion coefficients $c_1 \cdots c_6$ can be written as 

{\small
\begin{eqnarray}
E c_1 & = & H_1 c_1 + U_1 \sqrt{n} c_2 + U_1 \sqrt{n+1} c_3 + U_2 \sqrt{n} c_4 + U_2 \sqrt{n+1} c_5 \nonumber \\
E c_2 & = & H_2 c_2 + U_1 \sqrt{n} c_1 + U_2 \sqrt{n} c_6 \nonumber \\
E c_3 & = & H_3 c_3 + U_1 \sqrt{n+1} c_1 + U_2 \sqrt{n+1} c_6 \nonumber
\\
E c_4 & = & H_4 c_4 + U_2 \sqrt{n} c_1 + U_1 \sqrt{n} c_6 \nonumber \\
E c_5 & = & H_5 c_5 + U_2 \sqrt{n+1} c_1 + U_1 \sqrt{n+1} c_6 \nonumber
\\
E c_6 & = & H_6 c_6 + U_2 \sqrt{n} c_2 + U_2 \sqrt{n+1} c_3 + U_1 \sqrt{n} c_4 + U_1 \sqrt{n+1} c_5
\end{eqnarray}
}

\noindent
The algebraic elimination of the coefficients $c_2$ through $c_5$ produces two
coupled equations of the form
\begin{eqnarray}
E c_1 & = & [H_1 + \Sigma_1(E)] c_1 + V_{16}(E) c_6 \nonumber \\
E c_6 & = & [H_6 + \Sigma_6(E)] c_6 + V_{61}(E) c_1 
\end{eqnarray}

\noindent
where

{\small
\begin{equation}
V_{16}(E) 
~=~   
V_{61}(E) 
~=~ 
U_1 U_2 
\left[\frac{n}{E - H_2} + \frac{n+1}{E - H_3} + \frac{n}{E - H_4} + \frac{n+1}{E - H_5} \right] 
\label{V16}
\end{equation}
}

\noindent
The self-energy terms $\Sigma_1(E)$ and $\Sigma_6(E)$ are small in the case of weak coupling, and
contribute to the energy at which the resonance occurs, but not to the splitting.

\subsection{Level splitting}

Within this model, the resonance condition is

\begin{equation}
H_1 + \Sigma_1(E)
~=~
H_6 + \Sigma_6(E)
\end{equation}

\noindent 
which is the finite-basis approximation equivalent to Equation (\ref{ResCond}) when
no oscillator quanta are exchanged in association with the excitation transfer process.
The level splitting at resonance in this finite basis model is

\begin{equation}
\delta E_{min} ~=~ 2 \sqrt{ V_{16}(E) V_{61}(E) }
\end{equation}

\noindent
If the self-energy terms can be neglected, then the energy eigenvalue $E$ is very nearly
equal to the unperturbed level energy of the initial and final states

\begin{equation}
E ~=~ H_1 ~=~ H_6
\label{res}
\end{equation}

\noindent
We can evaluate the indirect coupling term $V_{16}(E)$ at resonance to be

\begin{equation}
V_{16}(E) 
~=~  
\frac{2 \hbar \omega_0 U_1 U_2}{\left(\Delta E_1\right)^2 -(\hbar \omega_0)^2} 
\label{V16_E}
\end{equation}

\noindent
where $\Delta E_1$ must approximately the same as $\Delta E_2$ for the resonance
condition of Equation (\ref{res}) to be satisfied.
When the characteristic oscillator energy $\hbar \omega_0$ is much smaller than
the transition energy $\Delta E_1$, then the level splitting on resonance 
$\delta E_{min}$ in this approximation evaluates to

\begin{equation}
\delta E_{min} ~=~ {4 \hbar \omega_0 g_1 g_2 \over n}
\end{equation}

\noindent
This is equivalent to the splitting we obtained using degenerate perturbation theory
in the rotated frame [Equation (\ref{eq:DeltaEmin})] as long as the integral $I_{12}$
is unity.
From an inspection of Equation (\ref{WKBapprox}), we see that this is the case as long
as 

\begin{equation}
g_1,g_2 ~\ll~1
\end{equation}

\noindent
Our result obtained in this section then is the weak coupling limit of what we obtained
above.
The generalization to a multi-mode system is considered in Appendix B.

\subsection{Destructive interference and loss}

The second-order coupling coefficient $V_{16}(E)$ for indirect coupling between state $\Phi_1$ and
$\Phi_6$ is much smaller than the direct coupling coefficients with intermediate states.
For example, we may write

\begin{equation}
{V_{16}(E) \over U_1 \sqrt{n}}
~\approx~
{2g_2 \over n} {\hbar \omega_0 \over \Delta E_1}
\end{equation}

\noindent
This reduction of coupling strength is due to destructive interference between the different
contributions in Equation (\ref{V16}) that make up $V_{16}(E)$.

To show that this is so, we consider a modified version of the model in which the destructive
interference is removed.
Consider the two-spin plus oscillator Hamiltonian augmented with a loss term $-i \hbar \hat{\Gamma}(E)/2$

$$
\hat{H}_{l} 
~=~  
  \Delta E_1 \frac{\hat{s}_z^{(1)}}{\hbar} 
+ \Delta E_2 \frac{\hat{s}_z^{(2)}}{\hbar} 
+ \hbar \omega_0 \hat{a}^\dagger \hat{a} 
- \frac{i \hbar}{2} \hat{\Gamma}(E)
\ \ \ \ \ \ \ \ \ \ \ \ \ \ \ \ \ \ \ \ \ \ \ 
\ \ \ \ \ \ \ \ \ \ \ \ \ \ \ \ \ \ \ \ \ \ \ 
$$
\begin{eqnarray}
\ \ \ \ \ \ \ \ \ \ \ \ \ \ \ \ \ \ \ \ \ \ \ 
+ U_1 \left(\hat{a}^\dagger + \hat{a} \right) \frac{2 \hat{s}_x^{(1)}}{\hbar} 
+ U_2 \left(\hat{a}^\dagger + \hat{a} \right) \frac{2 \hat{s}_x^{(2)}}{\hbar} 
\end{eqnarray}

%

\noindent
This loss term accounts for energetic decay processes of the low-lying intermediate states 
at the transition energy of the two-level systems.
Only intermediate states with both two-level systems in the ground state ($\Phi_2$ and $\Phi_3$) 
can decay in this model, since these states have unperturbed energies much less than the
available energy $E$.
The system cannot decay in such a way as to produce $\Phi_4$ and $\Phi_5$ as final states.

The coefficients $c_2$ and $c_3$ in this kind of model now satisfy

{\small

\begin{eqnarray}
E c_2 & = & \left(H_2 - \frac{i \hbar \Gamma}{2} \right) c_2 + U_1
\sqrt{n} c_1 + U_2 \sqrt{n} c_6 \nonumber \\
E c_3 & = & \left(H_3 - \frac{i \hbar \Gamma}{2} \right) c_3 + U_1 \sqrt{n+1}
c_1 + U_2 \sqrt{n+1} c_6
\end{eqnarray}

}

\noindent
The indirect coupling coefficient $V_{16}(E)$ in this model is changed due
to loss effects.
We may write

{\small

\begin{equation}
V_{16}(E) = U_1 U_2 \left[\frac{n}{E - H_2 + i \hbar \Gamma/2} +
  \frac{n+1}{E - H_3 + i \hbar \Gamma/2} + \frac{n}{E - H_4} +
  \frac{n+1}{E- H_5} \right]
\end{equation}

}

\noindent
In the limit that the loss term becomes very large 
then we may write 

\begin{equation}
V_{16}(E) 
~=~ 
-{
2 U_1 U_2 \Delta E_1 n 
\over
(\Delta E_1)^2 - (\hbar \omega_0)^2 
} 
\ \ \ \ \ \ \ \ \ \
(\Gamma \to \infty)
\label{V16loss}
\end{equation}

\noindent
The ratio of this indirect coupling coefficient (which is now free of 
destructive interference effects) to the direct coupling coefficient
between states $\Phi_1$ and $\Phi_2$ becomes

\begin{equation}
{V_{16}(E) \over U_1 \sqrt{n}}
~\approx~
-2 g_2
\end{equation}

\noindent
The removal of destructive interference increases the indirect coupling
coefficient by a factor of

\begin{equation}
{ 
|V_{16}(E)|_{\Gamma = \infty} 
\over 
|V_{16}(E)|_{\Gamma = 0}
} 
~=~
{n \Delta E_1 \over \hbar \omega_0}
\end{equation}


  In Appendix C, we consider a different modification of the model in which
the coupling between the different two-level systems and oscillator is through
conjugate oscillator operators.
Part of the interference is removed in this model.

\newpage
\section{Summary and conclusions}

In previous publications, we studied the shifts and splitting of energy levels
in the multiphoton regime in the spin-boson model, 
and in generalizations of the spin-boson model to the case of an oscillator coupled
to a spin-one system, and to a three-level system.
In these works, we made use of a unitary transformation which rotates the Hamiltonian
into a form in which terms primarily responsible for energy level shifts are
separated from terms responsible for the level splitting.
Here the same general approach is applied to a model involving a pair of two-level 
systems coupled to an oscillator, where the transition energies and coupling strengths
can be different.
The energy levels can be approximated accurately using a WKB approximation
based on the unperturbed part of the rotated Hamiltonian $\hat{H}_0$, similar to what we
found previously.
The level splitting at the Bloch-Siegert anticrossings are also described accurately
(away from anticrossing resonances) using degenerate perturbation theory based on
the $\hat{V}_1$ and $\hat{V}_2$; the situation is very similar to what we found
in the spin-boson problem.

What is new in this model (with no analog in the spin-boson problem) is the excitation transfer effect,
in which a transition in one two-level system occurs in concert with a transition in the other two-level system.
Approximate level energies derived from the unperturbed Hamiltonian $\hat{H}_0$ can
be used to locate excitation transfer resonances (see Figure \ref{Fig35x1}).
The excitation transfer effect is mediated by the spin-spin term $\hat{V}_{12}$ in the rotated Hamiltonian;
level splittings away from the Bloch-Siegert (energy exchange) resonances are accurately
modeled using degenerate perturbation theory based on the spin-spin term.
The effect is weak in this model due to destructive interference effects.
Destructive interference was examined in the context of a finite-basis approximation appropriate to
the weak coupling limit of the model, in which contributions from different paths 
can be seen to cancel.
A version of the model augmented with loss in the intermediate states removes the
destructive interference, and leads to a drastic increase in the indirect coupling
term.

A consequence of the destructive interference is that the excitation transfer effect is
independent of mode excitation in the weak coupling limit.
If there are many modes present, then it is likely that interference between the 
contributions of the different modes will further decrease the effect, especially
if the atoms or molecules represented by the two-level systems are far apart.
This effect is discussed in Appendix B.
Hence, one would expect excitation transfer to be a weak and short range effect in a
multi-mode system where the destructive interference involving different
pathways is not removed.
Such is the case 
in phonon-mediated excitation transfer.

One question raised from the analysis and discussion presented
here concerns the possibility of developing an excitation transfer system in
which the destructive interference is reduced or eliminated.
In this paper we have noted a reduction of interference in a modified version
of the model with conjugate coupling (Appendix C), and elimination in a model with large
loss in intermediate states (Section 6).
In either case, the excitation transfer rates and level splitting
will be greatly increased.
A conjugate coupling scheme could be developed in an electromagnetic resonator
under conditions where one two-level system couples to the electric field, and
the other two-level system couples to the magnetic field (since electric and
magnetic fields in a resonator are conjugate variables).
To implement a physical system in which loss impacts the destructive interference,
one would require a strongly driven mode which sees low loss at the
resonant frequency, but is very lossy at higher frequency corresponding to
the transition energies.

\newpage
\appendix

\section{WKB approximation}

We have found the WKB approximation to be very effective for the large $n$
limit of this problem.
Within the WKB approximation, the eigenfunctions $u(y)$ are written in
the form \cite{HagelsteinBook}

\begin{equation}
u(y) ~=~ C {\sin \phi \over \sqrt{\eta}}
\end{equation}

\noindent
where

\begin{equation}
{d \over dy} \phi(y) ~=~ \eta(y)
\end{equation}

\begin{equation}
\eta(y) ~=~ \sqrt{\epsilon - v(y)}
\end{equation}

\noindent
and where $C$ is a normalization constant.
Written in terms of WKB eigenfunctions, the integral $I_{12}$ becomes

\vskip 0.1in

{\small

\begin{equation}
I_{12}
~=~
C_0 C_1
\int_{-\infty}^\infty
{
\sin[\phi_0(y)] \sin[\phi_1(y)] 
\over
\sqrt{\eta_0(y) \eta_1(y)}
\left ( 1 + \displaystyle{8U_1^2 y^2 \over \Delta E_1^2} \right )
\left ( 1 + \displaystyle{8U_2^2y^2 \over \Delta E_2^2} \right )
}
dy
\end{equation}
}

\vskip 0.1in

The product of sine functions in the numerator can be decomposed into
terms which involve rapid oscillations, and slow oscillations.
For this, we can make use of the trigonometric identity

\begin{equation}
\sin(\phi_0) \sin(\phi_1) ~=~ {\cos(\phi_1-\phi_0) - \cos(\phi_1+\phi_0) \over 2}
\end{equation}

\noindent
On the RHS, the first cosine function has a slowly varying phase, and the second one has a
rapidly varying phase (especially if $n$ is very large).
Usually the contribution from the cosine with the slowly varying phase dominates, in
which case we may write

\vskip 0.2in

{\small

\begin{equation}
I_{12}
~=~
{C_0 C_1 \over 2}
\int_{-\infty}^\infty
{
\cos[\Delta \phi(y)]  
\over
\eta(y) 
\left ( 1 + \displaystyle{8U_1^2 y^2 \over \Delta E_1^2} \right )
\left ( 1 + \displaystyle{8U_2^2 y^2 \over \Delta E_2^2} \right )
}
dy
\end{equation}
}

\vskip 0.1in

\noindent
In writing this we assume that the two WKB momentum variables $\eta_0$ and $\eta_1$ are not very different

\begin{equation} 
\eta_0(y) ~\approx~ \eta_1(y) ~\to~ \eta(y)
\end{equation}

\noindent
The normalization constants $C_0$ and $C_1$ in this approximation satisfy

\begin{equation}
{C^2 \over 2} 
\int_{-\infty}^\infty
{1 \over \eta(y)}
dy
~=~
1
\end{equation}

\noindent
This allows us to recast the WKB approximation in the form

\vskip 0.1in

{\small

\begin{equation}
I_{12}
~=~
{
\displaystyle{
\int_{-\infty}^\infty
{
\displaystyle{\cos[\Delta \phi(y)]}  
\over
\displaystyle{\eta(y)} 
\left ( 1 + \displaystyle{8U_1^2 y^2 \over \Delta E_1^2} \right )
\left ( 1 + \displaystyle{8U_2^2 y^2 \over \Delta E_2^2} \right )
}
dy
}
\over
\displaystyle{
\int_{-\infty}^\infty
\displaystyle{1 \over \eta(y)}
dy
}
}
\label{WKBapprox}
\end{equation}

}
\vskip 0.1in

In Table \ref{tab35x1} we give the results of computations of the magnitude of
the integral $|I_{12}|$ from the numerical solution of Equation (\ref{Schrod_y}),
and from the WKB approximation of Equation (\ref{WKBapprox}).
One sees that the WKB approximation in this case is very close to the numerically
exact result.

\begin{table}[t]
\caption{Results for $I_{12}$ at resonance for $\Delta E_1 = 11 ~\hbar \omega_0$ and
$\Delta E_2 = 15 ~\hbar \omega_0$ for $n_0=9600$.
Values of the dimensionless coupling constants $g_1$ and $g_2$ at resonance (first
two columns);
integral computed from eigenfunctions of $\hat{H}_0$ (third column);
integral computed from the WKB approximation (fourth column).
}
\label{tab35x1}
\centering
\begin{tabular}{cccc}
 \cr
\hline
$g_1$ & $g_2$ & $|I_{12}|$ & $|I_{12}[WKB]|$  \cr
\hline
0.50 & 0.1611206  & 0.2224 & 0.2235 \cr
0.60 & 0.2734154  & 0.1946 & 0.1951 \cr
0.70 & 0.3666892  & 0.1733 & 0.1734 \cr
0.80 & 0.4528576  & 0.1562 & 0.1562 \cr
0.90 & 0.5353554  & 0.1421 & 0.1420 \cr
1.00 & 0.6156505  & 0.1303 & 0.1301 \cr
\hline\end{tabular}

\end{table}

\newpage
\section{Multi-mode case}

There is an additional mechanism that produces destructive interference in
the multi-mode generalization of this problem.
In particular, interference can occur between the contributions of the different
modes.

\subsection{Multi-mode Hamiltonian}

For simplicity, suppose that we have many modes that can be described in terms of
wavevectors ${\bf k}$; and that the mode frequencies are a function of the wavevectors
according to the dispersion relation

\begin{equation}
\omega ~=~ \omega({\bf k}) ~\to~\omega_{\bf k}
\end{equation}

\noindent
The multi-mode generalization of the two-spin problem then might be described
according to a Hamiltonian $\hat{H}_{mm}$ of the form

{\small

$$
\hat{H}_{mm} 
~=~  
\Delta E_1 {\hat{s}_z^{(1)} \over \hbar}
+
\Delta E_2 {\hat{s}_z^{(2)} \over \hbar}
+
\sum_{\bf k} \hbar \omega_{\bf k} \hat{a}^\dagger_{\bf k} \hat{a}_{\bf k}  
+ 
\sum_{\bf k} 
\bigg [
U_1({\bf k}) \hat{a}_{\bf k} e^{i {\bf k} \cdot {\bf r}_1} 
+
U_1^*({\bf k}) \hat{a}_{\bf k}^\dagger e^{-i {\bf k} \cdot {\bf r}_1} 
\bigg ]
{2 \hat{s}_x^{(1)} \over \hbar} 
$$
\begin{equation}
+ 
\sum_{\bf k} 
\bigg [
U_2({\bf k}) \hat{a}_{\bf k} e^{i {\bf k} \cdot {\bf r}_2} 
+
U_2^*({\bf k}) \hat{a}_{\bf k}^\dagger e^{-i {\bf k} \cdot {\bf r}_2} 
\bigg ]
{2 \hat{s}_x^{(2)} \over \hbar} 
\label{eq:TwoTwoLevelMultimodeHamiltonian}
\end{equation}

}

\noindent 
where ${\bf r}_1$ and ${\bf r}_2$ are the position vectors of the two atoms.

\subsection{Indirect coupling without loss}

It is possible to develop a finite-basis model similar to that discussed in Section 6, but
now for the multi-mode case, to obtain the indirect coupling term for the weak-coupling limit.
If we carry out such a calculation, in place of Equation (\ref{V16_E}), 
we obtain

\vskip 0.1in

{\small

$$
V_{16}(E) 
~=~
\frac{2 \hbar \omega_0 U_1 U_2}{\left(\Delta E_1\right)^2 -(\hbar \omega_0)^2} 
~\to~
\ \ \ \ \ \ \ \ \ \ \ \ \ \ \ \ \ \ \ \ \ \ \ 
\ \ \ \ \ \ \ \ \ \ \ \ \ \ \ \ \ \ \ \ \ \ \ 
\ \ \ \ \ \ \ \ \ \ \ \ \ \ \ \ \ \ \ \ \ \ \ 
$$
\begin{equation}
\int {U_1({\bf k}) U_2^*({\bf k}) e^{i {\bf k} \cdot ({\bf r}_1 - {\bf r}_2)} \over \Delta E_1 - \hbar \omega_{\bf k}} 
d^3{\bf k}
-
\int {U_1^*({\bf k}) U_2({\bf k}) e^{-i {\bf k} \cdot ({\bf r}_1 - {\bf r}_2)} \over \Delta E_1 + \hbar \omega_{\bf k}} 
d^3{\bf k}
\end{equation}

}

\vskip 0.1in

\noindent
The contribution of a single highly off-resonant mode appears on the same footing as contributions from
many other off-resonant modes, each weighted by a phase factor.
In the event that the separation $|{\bf r}_1-{\bf r}_2|$ is large, then severe cancelation can occur between
the contributions of the different modes; hence we would expect interactions to be of short range in
the multi-mode case.
Indirect coupling through coupling to common phonon modes would be described using such an approach (see for
example \cite{Vorrath2003}.
If one adopts for the multi-mode oscillator longitudinal photon modes (all with zero energy), then indirect
coupling between electric dipoles can be thought of in this way; with an overall $|{\bf r}_1-{\bf r}_2|^{-3}$ 
dependence of the dipole-dipole matrix element in the absence of retardation for single longitudinal photon
exchange.
Excitation transfer in this case is known as resonance energy transfer in biophysics \cite{Andrews}.

In general, any long range interactions in this kind of model will be dominated by resonant interactions
if resonant modes exist.  
Otherwise, the indirect interaction will most likely occur only at short range.

\subsection{Indirect coupling with loss}

If some mechanism is present that can remove the destructive interference effect in the different
pathways, then it may be possible to reduce or eliminate the destructive interference effect
in the multi-mode version of the problem.
The Hamiltonian in this case can be written as

{\small

$$
\hat{H}_{mml} 
~=~  
\Delta E_1 {\hat{s}_z^{(1)} \over \hbar}
+
\Delta E_2 {\hat{s}_z^{(2)} \over \hbar}
+
\sum_{\bf k} \hbar \omega_{\bf k} \hat{a}^\dagger_{\bf k} \hat{a}_{\bf k}  
- i {\hbar \hat{\Gamma}(E) \over 2}
+ 
\sum_{\bf k} 
\bigg [
U_1({\bf k}) \hat{a}_{\bf k} e^{i {\bf k} \cdot {\bf r}_1} 
+
U_1^*({\bf k}) \hat{a}_{\bf k}^\dagger e^{-i {\bf k} \cdot {\bf r}_1} 
\bigg ]
{2 \hat{s}_x^{(1)} \over \hbar} 
$$
\begin{equation}
+ 
\sum_{\bf k} 
\bigg [
U_2({\bf k}) \hat{a}_{\bf k} e^{i {\bf k} \cdot {\bf r}_2} 
+
U_2^*({\bf k}) \hat{a}_{\bf k}^\dagger e^{-i {\bf k} \cdot {\bf r}_2} 
\bigg ]
{2 \hat{s}_x^{(2)} \over \hbar} 
\label{eq:TwoTwoLevelMultimodeHamiltonian}
\end{equation}

}

\noindent
An analogous finite-basis type model can be developed for this kind of model,
leading to an indirect coupling in the $\Gamma \to \infty$ limit of

{\small

$$
V_{16}(E) 
~=~
-{
2 U_1 U_2 \Delta E_1 n 
\over
(\Delta E_1)^2 - (\hbar \omega_0)^2 
} 
~\to~
\ \ \ \ \ \ \ \ \ \ \ \ \ \ \ \ \ \ \ \ \ \ \ 
\ \ \ \ \ \ \ \ \ \ \ \ \ \ \ \ \ \ \ \ \ \ \ 
\ \ \ \ \ \ \ \ \ \ \ \ \ \ \ \ \ \ \ \ \ \ \ 
$$
\begin{equation}
- \int {U_1({\bf k}) U_2^*({\bf k}) n({\bf k}) e^{i {\bf k} \cdot ({\bf r}_1 - {\bf r}_2)} \over \Delta E_1 - \hbar \omega_{\bf k}} 
d^3{\bf k}
-
\int {U_1^*({\bf k}) U_2({\bf k}) [n({\bf k})+1] e^{-i {\bf k} \cdot ({\bf r}_1 - {\bf r}_2)} \over \Delta E_1 + \hbar \omega_{\bf k}} 
d^3{\bf k}
\end{equation}

}

\noindent
Things are qualitatively different in this case, since the indirect coupling now depends on the number of quanta $n({\bf k})$
in the different modes.
If a single off-resonant mode is very highly excited, then the interaction can take place over a much longer range since
the unexcited mode can no longer destructively interfere.

\newpage
\section{Model with conjugate couplings}

It is possible to eliminate some of the destructive interference by allowing the two different two-level systems
to couple with the oscillator through interactions that do not commute.
Consider a modified version of the model described by the Hamiltonian

{\small

\begin{equation}
\label{eq:TwoTwoLevelHamiltonian}
\hat{H}_{c} 
~=~
\Delta E_1 {\hat{s}_z^{(1)} \over \hbar}
+
\Delta E_2 {\hat{s}_z^{(2)} \over \hbar}
+
\hbar \omega_0  \hat{a}^\dagger \hat{a} 
+
U_1 (\hat{a}^\dagger+\hat{a}) {2 \hat{s}_x^{(1)} \over \hbar}
+i
U_2 (\hat{a}^\dagger-\hat{a}) {2 \hat{s}_x^{(2)} \over \hbar}
\end{equation}

}

\noindent
If adopt the same finite-basis approximation for the wavefunction $\Psi$ as discussed in Section 6

\begin{equation}
\Psi ~=~ \sum_{j=1}^6 ~c_j~\Phi_j
\end{equation}

\noindent
then the expansion coefficients satisfy

{\small
\begin{eqnarray}
E c_1 & = & H_1 c_1 + U_1 \sqrt{n  } c_2 +  U_1 \sqrt{n+1} c_3 + iU_2 \sqrt{n} c_4 - iU_2 \sqrt{n+1} c_5 \nonumber \\
E c_2 & = & H_2 c_2 + U_1 \sqrt{n  } c_1 + iU_2 \sqrt{n} c_6 \nonumber \\
E c_3 & = & H_3 c_3 + U_1 \sqrt{n+1} c_1 - iU_2 \sqrt{n+1} c_6 \nonumber
\\
E c_4 & = & H_4 c_4 - iU_2 \sqrt{n} c_1 + U_1 \sqrt{n} c_6 \nonumber \\
E c_5 & = & H_5 c_5 + iU_2 \sqrt{n+1} c_1 + U_1 \sqrt{n+1} c_6 \nonumber
\\
E c_6 & = & H_6 c_6 + iU_2 \sqrt{n} c_2 - iU_2 \sqrt{n+1} c_3 + U_1 \sqrt{n} c_4 + U_1 \sqrt{n+1} c_5
\end{eqnarray}
}

\noindent
The indirect coupling coefficient $V_{16}(E)$ in this case is

{\small

\begin{equation}
V_{16}(E) 
~=~
iU_1 U_2 
\left[ 
{n \over E - H_2} 
-
{n+1 \over E - H_3} 
+ 
{n \over E - H_4} 
-
{n+1 \over E- H_5} 
\right]
\end{equation}

}

\noindent
At resonance, we assume that

\begin{equation}
E ~=~ H_1 ~=~ H_6
\end{equation}

\noindent
to obtain

\begin{equation}
V_{16}(E) 
~=~ 
-{
2i U_1 U_2 \hbar \omega_0 (2n+1)
\over
(\Delta E_1)^2 - (\hbar \omega_0)^2
}
\end{equation}

\noindent
This indirect coupling is much larger than we obtained with the original model, 
and also depends on the mode excitation.
However, we see that it is smaller in magnitude by $\hbar \omega_0 / \Delta E_1$
than the result [Equation (\ref{V16loss})] we obtained in which the destructive interference was
completely eliminated by loss.
Hence, we have removed some of the destructive interference with conjugate coupling.

\end{document}